# Alpha rhythm slowing in temporal epilepsy across Scalp EEG and MEG

Vytene Janiukstyte[1], Csaba Kozma[1], Thomas W. Owen[1], Umair J Chaudhury[3], Beate Diehl[3], Louis Lemieux[3], John S Duncan[3], Fergus Rugg-Gunn[3], Jane de Tisi[3], Yujiang Wang[1,2,3], Peter N. Taylor[1,2,3,*]

1. CNNP Lab (www.cnnp-lab.com), Interdisciplinary Computing and Complex BioSystems Group, School of Computing, Newcastle University, Newcastle upon Tyne, United Kingdom

2. Faculty of Medical Sciences, Newcastle University, Newcastle upon Tyne, United Kingdom

3. UCL Queen Square Institute of Neurology, Queen Square, London, United Kingdom

* peter.taylor@newcastle.ac.uk

## Abstract

EEG slowing is reported in various neurological disorders including Alzheimer's, Parkinson's and Epilepsy. Here, we investigate alpha rhythm slowing in individuals with refractory temporal lobe epilepsy (TLE), compared to healthy controls, using scalp electroencephalography (EEG) and magnetoencephalography (MEG).

We retrospectively analysed data from 17,(46) healthy controls and 22,(24) individuals with TLE who underwent scalp EEG and (MEG) recordings as part of presurgical evaluation. Resting-state, eyes-closed recordings were source reconstructed using the standardized low-resolution brain electrographic tomography (sLORETA) method. We extracted low (slow) 6-9 Hz and high (fast) 10-11 Hz alpha relative band power and calculated the alpha power ratio by dividing low (slow) alpha by high (fast) alpha. This ratio was computed for all brain regions in all individuals.

Alpha oscillations were slower in individuals with TLE than controls ($p<0.05$). This effect was present in both the ipsilateral and contralateral hemispheres, and across widespread brain regions.

Alpha slowing in TLE was found in both EEG and MEG recordings. We interpret greater low (slow)-alpha as greater deviation from health.

## Introduction

The alpha rhythm 8-13Hz[1–5] is a dominant neurophysiological oscillation in the occipital and parietal regions[1,4,6], during a relaxed, eyes-closed, wakeful state observed in most humans[1,2,4]. It is believed to be generated by cortico-thalamic connections[7], and is alerted by: attention[1–3,8,9], language[3], sensory processing[1] and working memory[2,3,9]. Additionally, alpha may be involved in inhibitory mechanisms and suppression, which reduce cortical excitability in non-essential cortical regions at present[2,3,5,8–10]. Deficiency of the latter functions may alter the alpha rhythm and facilitate pathological activity[11–15]. Therefore, alpha rhythm alterations may indicate early brain health degradation.

EEG slowing is a hallmark of neurodegenerative, metabolic and neuropsychiatric diseases[3,16–23], which can be difficult to distinguish from healthy aging[21]. Alpha rhythm slowing has also been found in epilepsy, showing reduced alpha power and topographical shift to more frontal sites[19,20]. Recent observations supports alpha rhythm alterations[24,25], and reports reduced peak alpha frequency (PAF) in individuals with mesial temporal lobe epilepsy (TLE) and their asymptomatic relatives in EEG[26] and magnetoencephalography (MEG)[27,28]. Overall, alpha slowing in epilepsy remains under-investigated and its clinical significance is unknown.

The heterogeneity of alpha rhythm slowing across various diseases and their syndromes, is usually considered a physiological by-product. Specifically, the role of cortico-thalamic connections, cortical neural networks, and thalamic influences on alpha ocitations are heavily debated[1,2,7,9,15,17,29], with studies supporting the idea that alpha oscillations are generated, maintained, or propagated via multiple networks. Alternatively, the alpha rhythm is thought to be involved in the inhibitory mechanism, which is mediated by GABA interneurons. In this hypothesis, compromised GABA interneurons facilitate neuronal excitation, which relates to seizure generation[30,31] and is linked to Alzheimer's disease neuropathology[3,32–34]. Multiple genetic mutations associated with epilepsy syndromes[26,30,31,35], and antiseizure medication may also contribute to alpha power alterations[26,36]. Overall, a slower alpha rhythm is associated with behaviour symptoms that are common across multiple neurological diseases.

Here, we extend current literature, by extracting relative band power from scalp EEG and MEG individual adult cohorts. We also present the spatial shift across hemispheres in left and right TLE. First, we quantified the distribution of alpha power by computing low (slow)- and high (fast)-alpha in resting-state scalp EEG and MEG recordings, in healthy controls and individuals with TLE. Initially, we relate the loss of normalised power in scalp EEG and MEG recordings to previously published literature[24,27]. Next, we investigate the spatial patterns of alpha-power in individuals with TLE, compared to healthy controls.

## Methods

### Ethical approval

All methods were carried out in accordance with relevant guidelines and regulations, and all experimental protocols were approved by Newcastle University Ethics Committee (1804/2020).

### Scalp EEG and MEG subjects

EEG data were acquired from 17 healthy volunteers and 22 individuals with drug-resistant TLE undergoing presurgical evaluation at the National Hospital for Neurology and Neurosurgery (NHNN; part of the UCLH National Health Service Foundation Trust), Queen Square, London, U.K., were recruited. The patients had no previous neurosurgery or invasive neurosurgical procedures prior to the EEG recordings, which were collected for research purposes during pre-surgical assessment. These data have been described previously[12].

MEG data were collected from 46 healthy volunteers and 24 individuals with TLE who were undergoing pre-surgical evaluation.

There were no significant differences present between control and patient cohorts based on age and sex (Table 1). A summary of EEG and MEG cohorts is available in Supplementary Material 1 and 2.

*Table 1.* *Descriptive statistics of healthy controls and patients in scalp EEG and MEG cohorts.*

|  | Scalp EEG | | | MEG | | |
|---|---|---|---|---|---|---|
|  | Controls | Patients | Test statistic | Controls | Patients | Test statistic |
| **Number of subjects** | 17 | 22 | N/A | 46 | 24 | N/A |
| **Age (Y): Mean (SD)** | 31.9 (6.46) | 34.2 (10.1) | $p$=0.41, es=8.69 | 30.5 (6.7) | 33.9 (9.5) | $p$=0.08, es=-1.75 |
| **Sex: Male/Female (%)** | 11/6 (65%) | 9/13 (41%) | $p$=0.14, $\chi^2$=2.17 | 16/30 (35%) | 12/12 (50%) | $p$=0.48, $\chi^2$=1.52 |
| **Epilepsy lateralisation: Left/Right (%)** | N/A | 14/8 (64%) | N/A | N/A | 9/15 (36%) | N/A |
| **Age of epilepsy onset (Y): Mean (SD)** | N/A | 12.9 (8.3) | N/A | N/A | 14.2 (7.7) | N/A |
| **Duration of epilepsy (Y): Mean (SD)** | N/A | 21.3 (13.2) | N/A | N/A | 19.8 (11.8) | N/A |

## Scalp EEG recording and processing

Eyes-closed resting-state EEG data were recorded from 30 scalp electrodes using a commercial MR-compatible system (BrainAmp MR and Vision Analyzer) with a sampling rate of 5000 Hz using a common average reference, positioned according to the 10-20 internation system[37]. All EEG data were acquired at the MRI Unit of the Epilepsy Society (Chalfont St Peters, Buckinghamshire, UK). Ocular and cardiac activity was captured using two reference electrodes: EOG and ECG. The first 30 seconds of the recordings were disregarded to allow subjects to settle.

The EEG recordings were processed in MATLAB and Brainstorm, using previously described methods[12], as follows: EEG recordings were downsampled to 250 Hz. We used ICBM152 anatomical MRI template in standard space and the boundary element method (BEM) to create a realistic head model template. Digitised Brainstorm electrode locations were co-registered to the template, projecting electrodes perpendicularly to the nearest anatomical scalp location and manually reviewed to confirm satisfactory warping. We bandpass filtered the sensor time series between 1 and 47.5 Hz. We manually inspected the individual interictal recordings for artefacts, spikes or potential pathological events. Signal space projection (SSP) was performed to identify artefactual components (ECG heartbeat artefact) with manual intervention were necessary. Ocular artefacts could not be accurately identified and were therefore retained to preserve biological signal, particularly given that eyes-closed recordings were acquired.

The resulting data were then source reconstructed using the standardized low-resolution brain electrographic tomography (sLORETA) method with a realistic head model derived from BEM, and spatially down-sampled based on Lausanne parcellation, into 68 regions of interest (ROIs)[38]. The most artefact-free 60-second epoch segments were selected from each subject to represent normative power EEG baselines.

Note that in low-resolution parcellation some ROIs represent large neocortical areas and summate activity from multiple neighbouring locations. Constrained manifestation of source mapping orients the overlapping current magnitudes from contradictory sides of the sulci in opposite directions. To counteract these issues, the source recordings were sign-flipped and subsequently averaged to create a single time series per region.

## MEG recording and processing

Eyes-closed resting-state MEG recordings were acquired using two 275-channel CTF whole-head MEG systems, at different locations. The healthy control cohort data were collected at CUBRIC, Cardiff U.K. (as part of the MEG UK partnership), and the patients were recruited at

NHNN and the data recorded at the Wellcome Centre for Human Neuroimaging of the UCL Queen Square Institute of Neurology, London, U.K.. The MEG data were processed in Brainstorm using previously described methods as follows[13]. In summary, MEG sensor locations were coregistered to the individual's MRI scans using fiducial markers and manually reviewed to confirm satisfactory alignment. MEG data were downsampled to 600 Hz. Independent component analysis (ICA) was used to aid manually removing ocular and cardiac artefacts. Artefact-free MEG recordings were source reconstructed using minimum-norm imaging technique sLORETA[39], and an overlapping spheres head model. Similar to EEG (see previous section), the time-series were spacially down-sampled into 68 ROIs. Finally, the most artefact-free 70-second segments were extracted to produce normative power MEG baselines.

## Computing relative scalp EEG and MEG power

To construct power spectral densities (PSDs) illustrated in Fig. 1B, we used the Welch's method using a 2-second sliding window with 1-second overlap in each neocortical region. We then normalised the PSD to sum to one, thus computing relative power. This was performed for each subject for each modality (i.e. EEG and MEG control and patient cohorts (Fig. 1A, B).

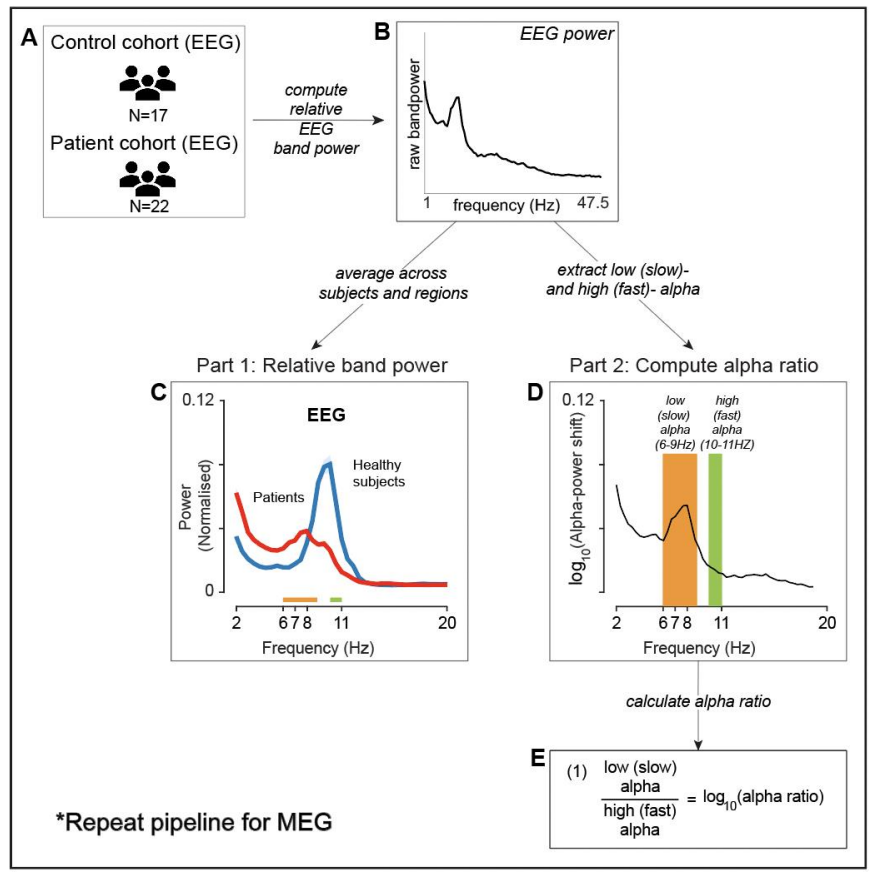

*Figure 1: Processing pipeline to calculate relative band power and to compute alpha power ratio in scalp EEG and MEG. (A) EEG recordings were collected for healthy controls and individuals with TLE cohorts. (B) We computed relative band power across individual subjects, for EEG recordings. (C) Relative band power was averaged across subjects and 68 ROIs and normalised. The resulting normalised power from 17 controls was plotted against 22 individuals with TLE, for EEG. (D) For each subject, we also extracted the low (slow)- 6 – 9 Hz and high (fast)- 10 – 11 Hz alpha relative band power, as described in[24]; (E) The 'alpha power log ratio' was then calculated as the ratio of low (slow)- over high (fast)- alpha power. *The processing pipeline is repeated for MEG data.*

### Plotting relative band power

To illustrate the alpha power shift, a measure of global band power was defined as the band power averaged across individuals and the 68 ROIs. We plotted the global band power for the healthy controls and patient cohorts band-wise between 1 and 20Hz, for EEG and MEG independently (Fig. 1C).

### Calculating alpha power log ratio

As a replication study, we had three aims: firstly, to validate the reported alpha shift from higher to lower frequencies, as previously reported[19,20,24,27]. Second, to investigate whether this effect was driven by hemispheric abnormality in TLE, and third, to evaluate whether the alpha rhythm shift was spatially widespread across the cortex.

To test our hypotheses, we focused on the low (slow)-alpha 6-9 Hz and high (fast)-alpha 10-11 Hz, as reported previously[24,40]. We defined our outcome variable as the "alpha-power ratio" by dividing low (slow)- alpha by high (fast)- alpha and taking the log function (Fig. 1D). The ratio expression simplifies the interpretation of the alpha power shift. We interpret high ratio score as having more low (slow)-alpha frequency, and we would expect the subject to deviate more from normality of healthy controls.

### Statistical analysis and data visualisation

We used the Anderson-Darling and Lilliefors tests to check the significance of deviations from normality in the EEG and MEG data. To compare the alpha power shift across subject cohorts we used unpaired t-tests. Similarly, we used paired t-tests to check whether spatially global alpha power shifts in controls vs patients were significant. Additional statistical tests with Bonferroni corrections are available in Supplementary Material 3 and 5. Visualization of the results was performed using the brainPlot[41] package implemented in Matlab.

## Results

Alpha rhythm slowing in TLE in scalp EEG and MEG recordings, replicates existing literature, shown in Fig. 2[24,27]. We also investigated the alpha power shift in individual patients (Fig. 3), and spatially, across hemispheres (Fig. 4). Each data point (alpha power ratio) is calculated by dividing low (slow)-alpha by high (fast)-alpha power and taking the log function.

### Alpha power shift

A shift in normalised alpha power to lower frequencies in EEG and MEG was observed in individuals with TLE compared to healthy controls, in agreement with previous findings (Fig. 2A, B).

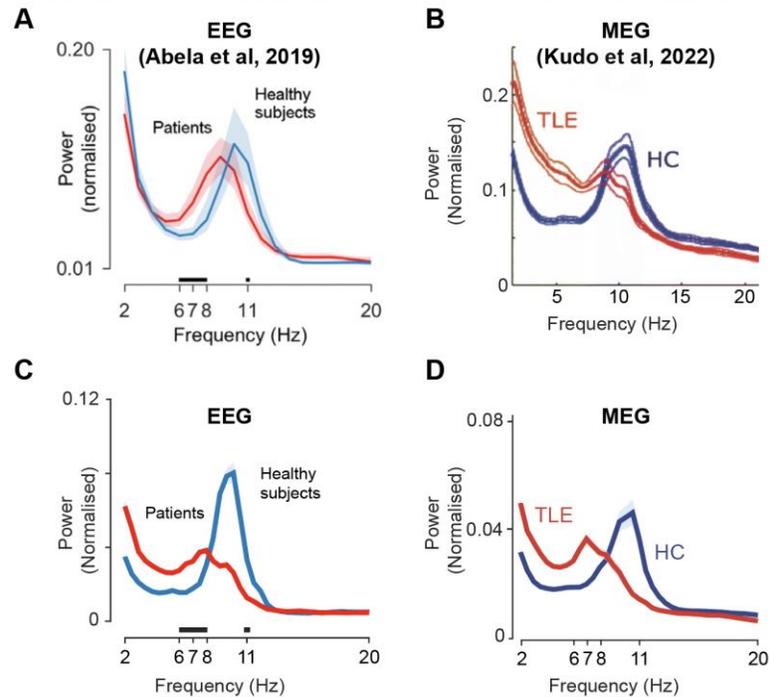

*Figure 2: Alpha shift in EEG and MEG. Alpha power peak is shifted towards lower-frequencies and reduced in TLE compared to healthy controls. Solid lines indicate the power at different frequency intervals and shaded areas illustrate 95% confidence intervals (CI). Plots (A and C) show the alpha shift in EEG and plots (B and D) show the power shift in MEG. The two plots in the upper row (A and B) illustrate previously published literature[24,27]. For visual aesthetics, colour and band power symbols were removed in plot (B). (see original in[27]. Both figures used with permission.*

## Alpha power shift in individual subjects

To quantify the previous observations, we compared data from individual subjects. Alpha power shift was present in patients with left TLE, in the EEG of both hemispheres, and independent of focus laterality and in both hemispheres in the MEG data (Fig. 3).

We used one-tailed t-tests to compare whether individual control and patient groups were significantly less or more than zero, respectively (Table 2). Both hemispheres in the left TLE patient cohort were significantly greater than zero (p=0.00, tstat=4.17, p=0.00, tstat=3.73) in EEG. In MEG, all control and patient cohorts were significantly different from 0 (p=0.00). Anderson-Darling and Lilliefors test of normality confirmed normality in EEG and MEG data.

In EEG, we found a significant bilateral alpha power ratio difference between control and left TLE patient group. Similarly, in MEG we found significant bilateral alpha power ratio difference between control groups and left and right TLE patient groups. The sample sizes for healthy control, left and right TLE patient cohorts in scalp EEG are: 17, 14 and 8, and in MEG: 46, 9 and 15 (per hemisphere). Additional statistical results are available in Supplementary Material 3.

*Table 2. Alpha power ratio tailed t-test results according to focus laterality and EEG/MEG hemispheres. P-values in bold remained significant after Bonferroni multiple comparison correction.*

| Subject cohort groups | Scalp EEG | | MEG | |
|---|---|---|---|---|
| | p | tstat | p | tstat |
| Healthy control left hemispheres < 0 | 0.05 | -1.74 | **0.00** | -4.20 |
| Healthy control right hemispheres < 0 | 0.04 | -1.92 | **0.00** | -4.12 |
| Left TLE patients left hemispheres > 0 | **0.00** | 4.17 | **0.00** | 3.89 |
| Left TLE patients right hemispheres > 0 | **0.00** | 3.73 | **0.00** | 4.00 |
| Right TLE patients left hemispheres > 0 | 0.14 | 1.16 | **0.00** | 4.55 |
| Right TLE patients right hemispheres > 0 | 0.12 | 1.31 | **0.00** | 4.40 |

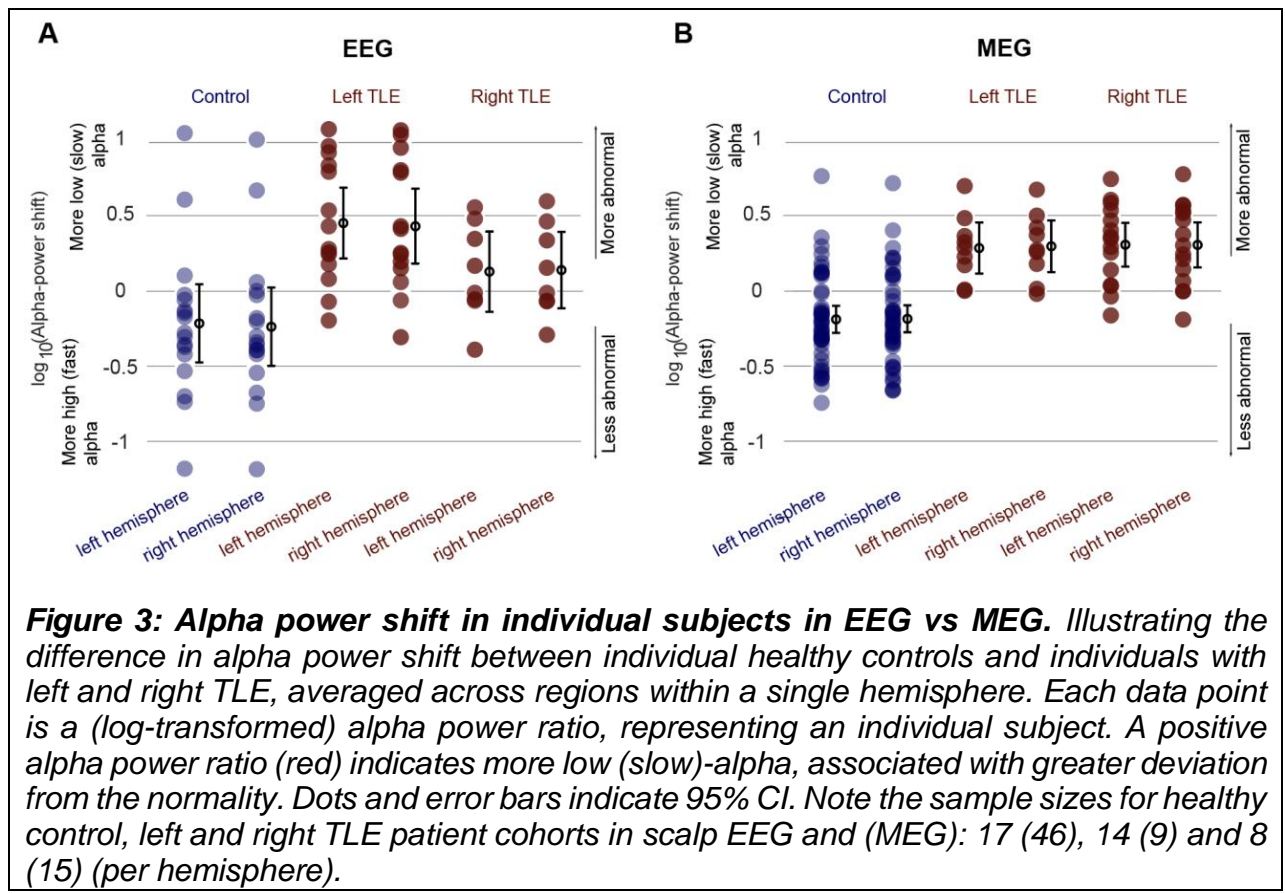

*Figure 3: Alpha power shift in individual subjects in EEG vs MEG. Illustrating the difference in alpha power shift between individual healthy controls and individuals with left and right TLE, averaged across regions within a single hemisphere. Each data point is a (log-transformed) alpha power ratio, representing an individual subject. A positive alpha power ratio (red) indicates more low (slow)-alpha, associated with greater deviation from the normality. Dots and error bars indicate 95% CI. Note the sample sizes for healthy control, left and right TLE patient cohorts in scalp EEG and (MEG): 17 (46), 14 (9) and 8 (15) (per hemisphere).*

## Spatial representation of alpha power shift across subjects in EEG and MEG modalities

Alpha power ratio brain maps are presented in Fig. 4. We report maximal low (slow) alpha in the anterior and temporal regions across most patient cohorts and in both modalities. The right hemisphere, in individuals with right TLE also shows low alpha spread in the occipital and parietal regions.

To quantify the spatial alpha shift between groups, we used paired t-tests, averaging the alpha power ratio across subjects, with each hemisphere comprising of 34 regions. After Bonferroni's correction, scalp EEG showed a high (fast)-alpha decrease across both patient hemispheres compared to controls, with more substantial differences in left TLE (p=0.00, tstat= -53.66 - -59.90). In MEG, there was a uniform loss of high (fast)-alpha across all TLE patient cohorts, compared to controls (statistical results are in Supplementary Material 5).

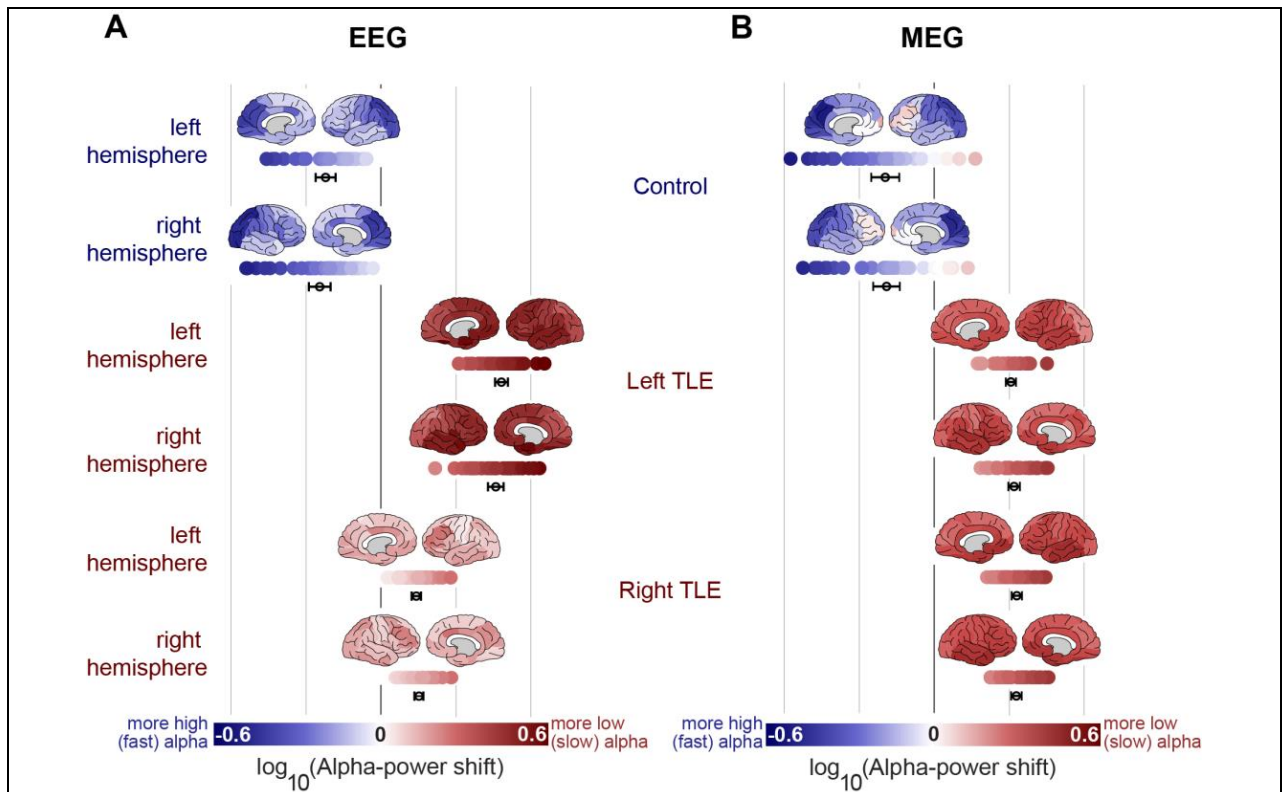

*Figure 4: Alpha shift is spatially global across patients.* Alpha-power shift varies spatially across the cortex in left and right TLE patients compared to healthy controls. Each data point is a (log-transformed) alpha ratio, representing a region of interest (ROI). There are 34 ROIs in each hemisphere. A positive (red) alpha ratio indicates more low (slow)-alpha, associated with greater deviation from health. Dots and error bars indicate 95% CI. (A and B) In the healthy control cohort, high (fast)-alpha is maximal in occipital and parietal regions, in EEG and MEG. Low (slow) alpha is maximal anteriorly and in the temporal regions, across most left and right TLE cohorts, in both modalities. Note that spatial locations of regions are for illustrative purposes.

## Discussion

Alpha power is shifted towards the lower frequencies in the TLE groups, across scalp EEG and MEG modalities, aligning with previous literature. Additionally, we illustrate the spatial alpha power shift from occipital to frontal regions. Taken together, our findings indicate that alpha rhythm slowing was found in individuals with TLE, specifically evident in the occipital and parietal regions where the higher frequency alpha predominates.

While a single routine EEG has less than 50% sensitivity to capture abnormal EEG waveforms[42], repeating routine EEGs increases detection of abnormality to over 90%[43,44]. We reaffirm the alpha power shift to the lower-frequencies in TLE, supporting previous literature[19,20,24,26–28]. We also demonstrate bilateral and widespread alpha slowing in TLE compared to controls, with a more pronounced effect seen in the left TLE on scalp EEG. Alpha slowing in MEG was significant for left and right TLE. Furthermore, we demonstrate the widespread spatial effect of slowing alpha rhythm (statistical results available in Supplementary Material 5).

The neurobiological explanation for the alpha rhythm slowing in epilepsy and other neurological conditions is unknown, although, large-scale networks are implicated[24,45]. For example, alpha oscillations are closely related to cortico-thalamic networks[1,2,8,9], and are known to propagate towards the occipital areas[2]. The thalamus is considered the peacemaker of alpha oscillations, having substantial control of the alpha rhythm[1,2,9]. Alpha rhythm has distinct spatial and functional patterns, suggesting multiple alpha generators with separate roles[7,9]. Specifically, the alpha rhythm is suspected to have divergent relationships with different parts of the default mode network (DMN)[46].

Sleep is a vital element of good health[15,47] and neurophysiologically involves multiple networks, such as the DMN[48]. Sleep deprivation is associated with alpha rhythm alterations[15], reducing functional connectivity within the DMN, and disrupted coupling within highly integrated and highly isolated networks[15,49]. GABA neurotransmitters regulate sleep patterns, and compromised function of GABAergic function impacts upon the quality of sleep[50]. Furthermore, the correlation between sleep deprivation and exacerbation of epileptic seizures is well acknowledged[24,51,52], and GABAergic dysfunction is hypothesised to be a major mechanism[53]. Similar findings have been reported in Alzheimer's disease, regarding alpha rhythm alterations[22], interrupted sleep[54] and disrupted GABA function[22,32,34]. Collectively, a pathological link may exist between epilepsy and Alzheimer's disease, presenting with similar symptoms and mechanisms.

This study had several strengths and limitations. One strength is the similarity of our results across two independent cohorts with different modalities, confirming published results. The subjects were approximately age- and sex-matched across both modalities. Additionally, eye-closure was enforced during data collection in scalp EEG and MEG. The limitations of this study include small sample sizes that narrows the variability pool, and statistical significance should therefore be interpreted with caution across groups. Potential age effects on band power were not considered, as our sample consisted of adult subjects only[55–57]. MEG data were collected at two different sites, which we did not control for. However, given that we were able to replicate previous results with data acquired in different sites, this suggests the alpha slowing effect is likely not driven by site differences. We also did not account for different epilepsy syndromes. The influence of daydreaming, circadian and ultradian rhythms on brain and bodily functions is widely acknowledged[58–60], however, their effects on cerebral functions and power remain a challenging area of research and were beyond the scope of this study. Furthermore, anti-seizure medication (ASM) are strong modulators of band power[26,36], which was not addressed in this study. Future studies could investigate the relationship between alpha rhythm changes and ASM usage. Lastly, a recent study found that asymptomatic relatives of individuals with epilepsy, not taking medication, also had reduced alpha frequency peak in the posterior regions and compromised somatosensory network function[26]. This suggests that alpha rhythm slowing may not directly relate to pathology.

In this replicatory study, we have shown significant alpha slowing in individuals with TLE compared to controls, and a forward spread of the low-frequency (slow) alpha, recorded with scalp EEG and MEG. Diverse neuropsychiatric diseases directly relate alpha rhythm alterations to behavioural symptoms. While we expand the epilepsy literature, we suggest that alpha slowing is a non-specific pathological symptom related to compromised neural networks, genetic mutations, and abnormal GABA expression. Future studies could explore whether alpha rhythm slowing could be reversed in people with epilepsy, and the implications of this.

## Acknowledgements

We thank members of the Computational Neurology, Neuroscience & Psychiatry Lab (www.cnnp-lab.com) for discussions on the analysis and manuscript; P.N.T. and Y.W. are both supported by UKRI Future Leaders Fellowships (MR/T04294X/1, MR/V026569/1). JSD, JdT are supported by the NIHR UCLH/UCL Biomedical Research Centre. The MEG normative data collection was supported by an MRC UK MEG Partnership Grant, MR/K005464/1.